**Control of Spatial-Temporal Congested Traffic Patterns at Highway Bottlenecks**

Boris S. Kerner

DaimlerChrysler AG, RIC/TS, HPC: T729
D-70546 Stuttgart, Germany
Phone: + 49 711 17 41453, Fax: +49 711 17 41242
boris.kerner@daimlerchrysler.com

**Abstract**

A microscopic theory of control of spatial-temporal congested traffic pattern at freeway bottlenecks is presented. Based on empirical spatial-temporal features of congested patterns at freeway bottlenecks which have recently been found, different control strategies for prevention or reducing of the patterns are simulated and compared. The studied control strategies include the on-ramp metering with feedback and automatic cruise control (ACC) vehicles. A recent microscopic traffic flow model within the author's three-phase traffic theory is used for validation of spatial-temporal congested pattern control.

# INTRODUCTION

Traffic management and control are some of the most important applications of traffic science. There are a huge number of publications and a lot of regular scientific conferences which are devoted to these subjects (e.g., *(1)-(5)*). One of the most efficient traffic pattern control methods is the ramp metering which is used on freeways in different countries (e.g., *(4)-(9)*). Another method which can be used for traffic control is application of automatic cruise control (ACC) vehicles (e, g., *(10)-(13)*). In all cases of effective traffic management and control of freeway traffic, a knowledge of *spatial temporal* congested patterns and their occurrence from a great importance.

In this article, a microscopic theory of control of spatial-temporal congested pattern at freeway bottlenecks is presented. This theory is made based on empirical spatial-temporal pattern features at freeway bottlenecks *(14)-(16)* and on microscopic traffic flow models *(17)-(19)* within the frame of the author's three-phase traffic theory *(20), (21)*. The control strategies include the on-ramp metering with feedback and automatic cruise control (ACC) vehicles. However before let us briefly discuss some empirical features of spatial-temporal congested patterns at freeway bottlenecks *(14)* and the microscopic model which will be used for simulations *(17), (18)*.

## Reproducible and Predictable Empirical Spatial-Temporal Pattern Features

Empirical studies of spatial-temporal congested patterns at freeway bottlenecks which have been made based on data measured during 1995-2003 have shown that the pattern possess a number of *reproducible and predictable spatial-temporal* features. This conclusion is also confirmed by the on-line application of the models "ASDA" and "FOTO" for spatial-temporal congested pattern recognition and tracing on different freeways in Germany *(22)-(24)*. These reproducible and predictable spatial-temporal pattern features are *(14)*:

(i) Types of congested patterns which spontaneous appear at a given effective freeway bottleneck or at a freeway section with several close to one another bottlenecks. This feature determines one of synchronized flow patterns (SP), or one of general patterns (GP), or else one of expanded congested patterns (EP) which occurs with *the highest probability* at a given traffic demand. Recall that a SP is a congested pattern which consists of the traffic phase "synchronized flow" only. There are three types of SPs: The moving SP (MSP), the widening SP (WSP) and the localized SP (LSP). GP is a congested pattern where synchronized flow occurs upstream of a freeway bottleneck and wide moving jams spontaneously emerge in that synchronized flow, i.e., the GP consists of the traffic phases "synchronized flow" and "wide moving jam". There are also different types of GPs. One of them is the dissolving GP (DGP) where only limited number of wide moving jams emerge. For GPs the case of "strong" congestion and "weak" congested should be distinguished. When two or more effective bottlenecks are close enough to one another on a freeway section then an expanded congested pattern (EP) can occur. In EP, synchronized flow region covers two or more effective locations of the bottlenecks. Wide moving jams which have occurred at a downstream bottleneck propagating through the upstream bottleneck where another moving jams are emerging (these wide moving jams are called "foreign" wide moving jams) can strong influence this moving jam emergence. These and other empirical features of spatial-temporal congested patterns at bottlenecks have been considered in *(14)* in detailed.

(ii) When traffic demand changes over time, then some certain type either of the pattern evolution or transformation between different congested patterns occurs with the highest probability.

(iii) Based on traffic measurements made on different days, the probability distribution of the certain pattern occurrence depending on traffic demand can be found. This gives the probability of the type of the congested pattern at a given effective freeway bottleneck or at a freeway section with several close to one another bottlenecks.

(iv) If this pattern is GP, then such characteristics of GP as whether weak congestion or strong congestion occurs in GP as well the mean length of the pinch region in GP, the average speed and the density in the pinch region can be found which the GP should have with the highest probability at a certain bottleneck.

(v) If this pattern is EP, then characteristics of EP such as whether there is only one or several separated pinch regions in EP, whether the "foreign" wide moving jam propagation is expected or it is not, as well as the approximate location of the upstream front of the most-upstream pinch region in EP can be predicted.

(vi) If there are bottlenecks where moving jams dissolve, then special types of congested patterns like GP where the region of wide moving jams is either fully or partially dissolving (this pattern can resemble LSP) can be predicted for some freeway sections.

Thus, based on measurements on different days of these *spatial-temporal reproducible and predictable* congested pattern features we can create a data bank for spatial-temporal congested pattern *(25)*. In this data bank, each set of spatial-temporal pattern features is related to the day which has the same characteristic (like weekend day, working day, events which are relevant for traffic and so on). If there is several different sets of spatial-temporal pattern features for the days with the same characteristics one should find a probability for each of such spatial-temporal pattern features. This permits the determination of the probability distribution for spatial-temporal pattern features.

If current measurements are available, then the following matching of the data of such data bank with current measurements may be used for the reliable prediction of spatial-temporal congested pattern features for a given freeway section. The spatial-temporal congested pattern characteristics can be found based on the FCD-technology *(26)*. In particular, a FCD-vehicle can determine the positions of the downstream front of a congested pattern at an effective bottleneck. Recall that within the downstream front of the congested pattern (WSP, LSP, GP or EP) the vehicle escape from the traffic phase "synchronized flow" to the traffic phase "free flow" *(14)*.

**About Microscopic Model within Three-Phase Traffic Theory used for Simulations**

The microscopic model within the three-phase traffic theory which is used in all simulations below is based on the following general rules of the vehicle motion introduced by Kerner and Klenov *(17)*:

$$v_{n+1} = \max(0, \min(v_{free}, v_{c,n}, v_{s,n})), \tag{1}$$

$$x_{n+1} = x_n + v_{n+1}\tau, \tag{2}$$

$$v_{c,n} = \begin{cases} v_n + \Delta_n & \text{at } x_{\ell,n} - x_n \leq D_n \\ v_n + a_n & \text{at } x_{\ell,n} - x_n > D_n \end{cases}, \tag{3}$$

where $v_n$ and $x_n$ are the speed and the space co-ordinate of the vehicle; the index $n$ corresponds to discrete time $t = n\tau$, $n = 0,1,2,...,$; $\tau$ is the time step; $v_{free}$ is the maximum speed in free flow which is considered as a constant value; $v_{s,n}$ is the save speed; $v_{c,n}$ (3) is a desirable speed; $\Delta_n$ is given by the formula:

$$\Delta_n = \max(-b_n\tau, \min(a_n\tau, v_{\ell,n} - v_n)), \tag{4}$$

$a_n$ is the vehicle acceleration, $b_n$ is the vehicle deceleration; the lower index $\ell$ marks variables related to the vehicle in front of the vehicle at $x_n$, the "leading vehicle"; all vehicles have the same length d; $D_n$ is "the synchronization distance": At

$$x_{\ell,n} - x_n \leq D_n \tag{5}$$

the vehicle tends to adjust its speed to the speed of the leading vehicle, i.e., the vehicle decelerates if $v_n > v_{\ell,n}$, and the vehicle accelerates if $v_n < v_{\ell,n}$. The synchronization distance $D_n$ as it has been shown in *(17), (19)* is related to the fundamental hypothesis of the author's three-phase traffic theory *(20), (21)*: Hypothetical steady states of synchronized flow (a steady state is the hypothetical model state where all vehicles move at the same distances to one another and with the same time-independent speed) cover a *two-dimensional* region in the flow-density plane (dashed region in Fig. 1 (a)). Thus, there is *no* fundamental diagram (*no* flow-density relationship) for steady model states in the model in accordance with the three-phase traffic theory.

Different possible descriptions of the functions $D_n$, $v_{s,n}$, model fluctuations, and delay times in the vehicle acceleration $a_n$ and deceleration $b_n$ which can be found in *(17)-(19)* lead to the same diagram of congested patterns (Fig. 1 (b)) which has first been postulated in the three-phase traffic theory *(14),(15)*, e.g., for spatial-temporal congested patterns occurred at a freeway bottleneck due to the on-ramp (Fig. 1 (c)).

**PATTERN CHARACTERISTICS UNDER MODULATED ON-RAMP INFLOW**

**On-Ramp Metering without Feedback**

Let us consider congested pattern control at an isolated freeway bottleneck due to an on-ramp (Fig. 1 (c)) which is at the freeway location $x = 16$ km on Fig. 2 (a-d). For all results presented below we will use the microscopic traffic flow model within the frame of the three-phase traffic theory *(17)* where the general rule of vehicle motion are (1)-(5). Corresponding to *(14), (18)*, there is a region of the flow rates on the main road upstream of the on-ramp $q_{in}$ and the flow rates to the on-ramp $q_{on}$ (Fig. 1 (c)) where GP occurs at the bottleneck in the diagram of congested patterns (Fig. 1 (b)). One of the possible GPs is shown in Fig. 2 (a). In this case, there is no on-ramp metering, i.e., the flow rate $q_{on}$ does not depend on time. Due to congestion upstream of the bottleneck the discharge flow rate, i.e., the flow rate in the outflow from a congested pattern at the bottleneck in free flow downstream of the bottleneck, $q_{out}^{(bottle)}$ (Fig. 2(e), curve 1) is noticeably lower than

$$q_{sum} = q_{in} + q_{on}. \qquad (6)$$

Now we consider the influence of the on-ramp control (the ramp metering) on congestion at the bottleneck. This control is applied when the vehicle speed on the main road in the on-ramp vicinity (at x=16 km) decreases below some given speed (85 km/h) which is considered as *the congestion criterion* (in model simulations, the free flow speed is 108 km/h). In this on-ramp control, if the speed is below the congestion criterion, then the flow rate $q_{on}$ due to a light signal on the on-ramp lane is periodically switched on during the green phase $T_G$ of the light signal and the flow rate to the on-ramp is zero during the red phase $T_R$. During this modulation of the on-ramp inflow the flow rate $q_{in}$ does not change. Then the average flow rate to the on-ramp $\bar{q}_{on}$ decreases:

$$\bar{q}_{on} < q_{on}. \qquad (7)$$

Note that although the average flow rate $\bar{q}_{on}$ is lower than $q_{on}$, at least at the beginning of the green phase the flow rate of vehicles squeezing from the on-ramp to the main road (Fig. 2 (g), marked as "16 km (on-ramp)") can be higher than $q_{on}$. This is because during the red phase a queue of vehicles is formed upstream of the location of the light signal on the on-ramp lane. The flow rate on the on-ramp lane in the outflow from the queue during the green signal is usually considerably higher than $q_{on}$. For this reason, in all cases considered below it occurs that $\bar{q}_{on} > q_{on} T_G /(T_G + T_R)$.

It is obviously that congestion at the bottleneck should decrease under the condition (7). Indeed, we find that rather than the initial GP a localized SP (LSP) occurs at the bottleneck (Fig. 2(b)). As a result of the decrease in congestion, the average discharge flow rate $q_{out}^{(bottle)}$ increases (Fig. 2(e), curve 2). This increase in $q_{out}^{(bottle)}$ occurs although

$$\overline{q}_{sum} = \overline{q}_{on} + q_{in} < q_{on} + q_{in}. \qquad (8)$$

**On-Ramp Metering with Feedback**

The on-ramp inflow control has a much more effect on the pattern dissolution and on the increase in the discharge flow rate $q_{out}^{(bottle)}$ if the on-ramp inflow control with a feedback is applied (Fig. 2 (c, d); $q_{out}^{(bottle)}$ =2060 in (a), 2173 in (b), 2177 in (c) and 2198 in (d) vehicles/h). In this case, as it has been made above the interruption of the flow rate to the on-ramp during the red phase is also applied when the vehicle speed on the main road in the on-ramp vicinity ( $x =16$ km ) decreases below the congestion criterion. However, the duration of the red phase $T_R$ is now changing over time in the range $[T_{R,1}, T_{R,2}]$. The current applied duration $T_R$ is determined by a feedback: Whether the speed on the main road is also lower than the congestion criterion at *a chosen distance* $\Delta x = 0.2$ km upstream of the merging region of the on-ramp $x =16$ km or it is not. If the congestion criterion at the distance $\Delta x$ is fulfilled, then $T_R$ reaches its maximum, $T_{R,2}$ In contrast, if the congestion criterion is only fulfilled in the on-ramp vicinity $x =16$ km but it is not fulfilled at the distance $\Delta x$ upstream then $T_R$ is set to the minimum, $T_{R,1}$.

Thus, if a congested pattern occurs, then we choose the control strategy using the feedback. This feedback allows us to localize the congested pattern within a small distance $\Delta x$. Owing to this control strategy, GP and WSP should dissolve, i.e., only a LSP can remain at the on-ramp. This indeed occurs (Fig. 2 (c, d)). In comparison with the periodical change in the green and red phases on the on-ramp, i.e., without the feedback (Fig. 2 (b)), the control strategy with the feedback increases both the discharge flow rate $q_{out}^{(bottle)}$ and the average speed inside congestion upstream of the bottleneck.

These increase in the discharge flow rate $q_{out}^{(bottle)}$ and the increase in the average speed in the congested pattern at the on-ramp depend on values $T_{R,2}$ and $T_{R,1}$. In particular, at lower $T_{R,1}$ we find a more increase in $q_{out}^{(bottle)}$ (Fig. 2 (c, d)). As it has already been mentioned, $q_{out}^{(bottle)}$ without the on-ramp control (curve 1 in Fig. 2 (e)) is considerably lower than with the on-ramp control (curve 2). The feedback in addition increases $q_{out}^{(bottle)}$ (curve 3 in Fig. 2 (e)). When the strategy with feedback is used, $q_{out}^{(bottle)}$ becomes a more smooth function of time (curve 3 in Fig. 2 (e)). Besides, the feedback allows us to decrease the duration of the minimum value $T_{R,1}$ of the red phase without a considerable decrease in the discharge flow rate (Fig. 2 (f), curve 4 - $q_{out}^{(bottle)}$ without feedback as function of $T_R$ ; curve 5 - $q_{out}^{(bottle)}$ with feedback as function of $T_{R,1}$ ). Thus, the on-ramp control strategy with feedback based on the traffic measurements upstream of a freeway bottleneck makes a considerable advantage for an increase in the flow rate on a freeway and for a decrease in traffic congestion at the freeway bottleneck and an increase in the speed in the pattern. The latter leads to a decrease in the travel time.

**DISSOLVING OF CONGESTED PATTERNS**

Let us a congested pattern has already occurred at the bottleneck. Then a decrease in the flow rate $q_{in}$ or in the flow rate $q_{on}$ can obviously lead to the dissolving of the pattern. However, the influence of these two flow rates occurs to be very *different* for the pattern dissolution. A study of this effect is the aim of this section.

Let us a GP under strong congestion *(14), (18)* has occurred at a freeway bottleneck due to the on-ramp at the initial flow rates $q_{in} = 2000$ vehicles/h and $q_{on} = 750$ vehicles/h (Fig. 3 (a)). Further we suggest that there is a detector 2.5 km upstream of the on-ramp. After the first moving jam in the forming GP has been recognized at this detector, the initial flow rate $q_{in}$, is decreased to some value $q_{in\,2} < q_{in}$ (Fig. 3 (b-f) $q_{in\ 2}$ = 1580 (b), 1530 (c), 1475 (d), 1150 (e), 950 (f) vehicles/h). We can see that the decrease in the flow rate $q_{in}$ however does not lead to the GP dissolving if the flow rate $q_{in}$ remains to be higher than the limit flow rate in the pinch region of the GP, $q_{lim}^{(pinch)}$ *(14), (18)* (Fig. 3 (b-d)). For the chosen model parameters $q_{lim}^{(pinch)} \approx 1200$ vehicles/h. Only when

$$q_{in} < q_{lim}^{(pinch)} \tag{9}$$

the GP dissolves (Fig. 3 (e, f)). This result means that a relative large decrease in $q_{in}$ is necessary for the GP dissolving.

The physical meaning of this result is linked to the nature of the pinch effect and the wide moving jam emergence in the GP *(14)*. The average flow rate in the pinch region of the GP under strong congestion is equal to $q_{lim}^{(pinch)}$ *(14)*. This flow rate is lower than the maximum flow rate in the wide moving jam outflow $q_{out}$ *(14), (18)*. If the flow rate $q_{in}$ becomes lower than $q_{out}$, then the most upstream wide moving jams in the GP dissolve first (Fig. 3 (b-d)). This wide moving jam has however almost no influence on the pinch region of the GP where narrow moving jams continuously remain to emerge even at

$$q_{in} < q_{out} \tag{10}$$

if the condition

$$q_{in} \geq q_{lim}^{(pinch)} \tag{11}$$

is fulfilled. Only when the opposite condition (9) is valid, the GP dissolves *(14), (18)*.

A more efficient dissolving of the GP can be made due to the on-ramp control. In Fig. 3 (g, h) the on-ramp control with the periodic green and red phases is applied ($T_G$ =20, $T_R$ =95 sec). We see that a decrease in the average flow rate $\bar{q}_{on}$ from 750 to 315 vehicles/h leads to the dissolving of GP even then if the flow rate $q_{in}$ does not change (Fig. 3 (g)). If, in addition, the initial flow rate $q_{in}$ is decreased to $q_{in\ 2}$ =1650 vehicles/h, then the congested pattern fully dissolves at the bottleneck (Fig. 3 (h)) although the flow rate $q_{in\ 2}$ still satisfies the condition (11).

The dissolving of WSP at the bottleneck can be reached by a lower decrease in $q_{in}$ even then if there is no on-ramp control (Fig. 4). When the flow rate $q_{in}$ is decreased, the average vehicle speed increases inside the WSP and the WSP upstream propagation first becomes slower and then it is interrupted up to the WSP fully dissolving.

**PREVENTION OF INDUCED CONGESTION AT UPSTREAM BOTTLENECK**

We have seen above that the dissolving of only wide moving jams inside the GP (Fig. 2 (d)) does not solve the problem of the congestion condition at the bottleneck.

Let us now consider two different bottlenecks, the "downstream" bottleneck and the "upstream" bottleneck. In Fig. 5, these bottlenecks are due to two spatially separated on-ramps: The downstream on-ramp 'D' is at $x_{on}^{(down)}$ =16 km and the upstream on-ramp 'U' is at $x_{on}^{(up)}$ =10 km. In this case, a dissolving of wide moving jams inside the GP which has occurred at the on-ramp 'D' can be very important for the prevention of the occurrence of the congested pattern at the on-ramp 'U'. Indeed, in empirical observations *(14)*, if a wide moving jam due to the jam upstream propagation reaches the upstream bottleneck, then the F $\rightarrow$ S transition (the breakdown phenomenon) can be *induced* at this bottleneck even then if free flow has been before at the upstream bottleneck.

This *induced* F → S transition leads to the congested pattern emergence at the upstream bottleneck. This induced speed breakdown is possible because the F → S transition is the first order phase transition: In free flow at the bottleneck a nucleation of the F → S transition is possible *(14)*. This nucleation is induced by the wide moving jam propagation. An example of the numerical simulation of such induced pattern emergence is shown in Fig. 5 (a): A wide moving jam from the GP at the on-ramp 'D' induces another GP at the on-ramp 'U'.

To dissolve both GP in Fig. 5 (a), the on-ramp metering control can be applied (Fig. 5 (b)). In Fig. 5 (b), after the first wide moving jam has been detected 2.5 km upstream of the on-ramp 'U', the flow rates both to the on-ramp 'U', $q_{on}^{(up)}$ and to the on-ramp 'D', $q_{on}^{(down)}$ are decreased due to the periodic switching of the light signals on the related on-ramp lanes. In Fig. 5 ($q_{in}$, $q_{on}^{(up)}$, $q_{on}^{(down)}$) are (1500, 500, 580) (a, b), (1800, 440, 30) (c, d), (1756, 480, 190) (e, f), (1978, 220, 180) (g, h) vehicles/h; $T_G$ =15 sec (b, d, f, h); $T_R$ =120 (b), 100 (d), 110 (f), 220 (h) sec; $\overline{q}_{on}^{(up)}$ = 217 (b), 236 (d), and 116 (h) vehicles/h.

Other examples of the dissolving of initial patterns in Fig. 5 (c, e, g) are presented in Fig. 5 (d, f, h), respectively. Figs. 5 (c, e, g) are related to the induced pattern formation at the upstream on-ramp 'U' due to the catch effect *(14)*. In this case, first a SP occurs at the downstream bottleneck. If the upstream front of synchronized flow in this pattern reaches the upstream bottleneck this front is caught at the bottleneck. As a result, a congested pattern can be formed at the upstream bottleneck. In Fig. 5 (c, e, g), MSP and two different WSP which have emerged at the on-ramp 'D' cause the induced SP pattern emergence at the on-ramp 'U', respectively. In the cases shown in Fig. 5 (e, g), EP emerge which synchronized flow regions cover both the downstream and upstream bottlenecks. To dissolve the congested patterns in Fig. 5 (c, e, g), after the formation of congestion has been detected 2.5 km upstream of the on-ramp 'U', the on-ramp metering control is used at the on-ramp 'U' only (Fig. 5 (d, f, h)).

**INFLUENCE OF AUTOMATIC CRUISE CONTROL ON CONGESTED PATTERNS**

In real ACC-systems (e.g., *(10)-(13)*), two or more ranges of the vehicle speed can usually be chosen where different dynamical rules for the ACC-vehicle are used. For a simplification, we will consider a hypothetical ACC system where there is only one dynamical rule for the ACC-vehicle in the whole possible range of the vehicle speed. In most known ACC-systems, at least in one of the speed ranges the dynamical behavior of the ACC-vehicle can approximately be described by the well-known equation (e.g., *(13)*):

$$a_n^{(ACC)} = K_1(g_n - \theta_0 v_n) + K_2(v_{\ell,n} - v_n), \qquad (12)$$

where $a_n^{(ACC)}$ is the acceleration or the deceleration of the ACC-vehicle at the time step n, $\theta_0$ is a desirable time gap which is a given parameter of the ACC-vehicle (below $\theta_0$ =1.8 sec). In (12), as well in the model (1)-(5) we will use the discrete time $t = n\tau$. The desirable time gap $\theta_0$ is usually set by the driver of the ACC-vehicle. $K_1$ and $K_2$ are the coefficients of the ACC adaptation. These coefficients describe the dynamical adaptation of the ACC-vehicle when either the space gap $g_n$ is different from $g_n = \theta_0 v_n$ or if the vehicle speed $v_n$ is different from the speed of the vehicle ahead $v_{\ell,n}$. If in contrast, $v_n = v_{\ell,n}$ and the condition $g_n = \theta_0 v_n$ is fulfilled then the ACC-vehicle acceleration and deceleration are zero, $a_n^{(ACC)} = 0$, i.e., the ACC-vehicle moves with a time-independent speed.

In a microscopic simulation model which is used to find an influence of the ACC-vehicles on spatial-temporal behavior of traffic at bottlenecks, there are vehicles which have no ACC-system and ACC-vehicles. The vehicles which have no ACC-system move in accordance with the microscopic model (1)-(5) within the three-phase traffic theory *(17)*. The ACC-vehicles are randomly distributed on the road between other vehicles which have no ACC-system. The ACC-vehicles move in accordance with (12). The maximal acceleration and the maximal deceleration of the ACC-vehicles have been limited by $2 \text{ m}/\text{s}^2$. The maximal speed of the ACC-vehicles is also limited either by the speed in free flow or a safe speed (these characteristic vehicle speeds are chosen as well in the model in (1)-(5)). The safe speed allows us to avoid collisions between vehicles.

Different percentages $\gamma$ of random distributed ACC vehicles between vehicles which have no ACC-system are investigated in different traffic situations. First possible influences of ACC-vehicles on initial congested traffic patterns at freeway bottlenecks and second the influence of the ACC-vehicles on initial free flow conditions are considered.

**Influence of ACC-vehicles on congested patterns**

To study the influence of ACC-vehicles on congested patterns, we consider first the flow rates $q_{in}$ and $q_{on}$ when GP occurs at the bottleneck due to the on-ramp when there are no ACC-vehicles on the freeway, i.e., at $\gamma = 0$ (Fig. 6 (a)).

ACC-vehicles can decrease the amplitude of moving jams in the initial GP, if the coefficients of the ACC-adaptation $K_1$ and $K_2$ in the dynamical equation of the ACC-vehicle are high enough. In this case, the ACC-vehicle quickly reacts on the changes in the gap and in the speed of the vehicle ahead. The suppression of moving jams in the initial GP due to the ACC-vehicles can be seen in Fig. 6 (b-d).

In Figs. 6 (b-d), the same $q_{in}$ and $q_{on}$ as well in Fig. 6 (a) are used. However, the percentage of the ACC-vehicles in traffic flow, $\gamma$, is increased from $\gamma = 20\%$ in Fig. 6 (b) to $\gamma = 37\%$ in Fig. 6 (d). There is some critical percentage of the ACC-vehicles $\gamma_{cr}$ (at chosen model parameters $\gamma_{cr} = 35\%$): When this critical percentage of the ACC-vehicles in traffic flow is reached, then there is almost no moving jams in the congested pattern upstream of the on-ramp any more, i.e., the initial GP in Fig. 6 (a) transforms into SP (Fig. 6 (d)). If percentage of the ACC-vehicles is further increased then no new moving jams occur in the SP. Thus, the ACC-vehicles can prevent the moving jam emergence.

However, the discharge flow rate $q_{out}^{(bottle)}$ is a continuously decreasing function of the percentage of the ACC-vehicles, $\gamma$ (Fig. 6 (e)). This is because the vehicle speed upstream of the on-ramp in the developed SP decreases when $\gamma$ increases. As a result, the flow rate inside the SP upstream of the on-ramp, $q_{con}$ also decreases (Fig. 6 (f)). Besides, due to the decrease in the speed in the SP with the increase in $\gamma$ the travel time for individual vehicles increases.

Thus, the ACC-vehicles with high enough coefficients of the ACC-adaptation prevent the moving jam emergence, i.e., a more comfortable and safety driving with the ACC-vehicles is possible. However, these ACC-vehicles can not prevent or dissolve traffic congestion. They lead even to some decrease in the discharge flow rate $q_{out}^{(bottle)}$ from the congested pattern at the bottleneck and to a travel time increase.

**Influence of ACC-vehicles on initially free traffic flow**

Another case occurs when the ACC-vehicle too slow reacts on the changes in the gap and in the speed of the vehicle ahead, i.e., the coefficients of the ACC-adaptation $K_1$ and $K_2$ in the dynamical equation of the ACC-vehicles (12) are low enough. Then the ACC-vehicles can lead to the onset of traffic congestion at the bottleneck even then, if there is no congestion in traffic flow without ACC-vehicles.

In Fig. 7 (a), where there is no ACC-vehicles, free flow occurs at the bottleneck at the chosen flow rates $q_{in}$ and $q_{on}$. There is some critical percentage of the ACC-vehicles, $\gamma_{cr}$ (at the chosen model parameters, $\gamma_{cr} = 4\%$): At the same $q_{in}$ and $q_{on}$, if ACC-vehicles appear on the freeway, then at low enough percentage of the ACC-vehicles

$$\gamma < \gamma_{cr} \qquad\qquad\qquad\qquad\qquad\qquad\qquad\qquad\qquad\qquad\qquad\qquad (13)$$

there is no influence of the ACC-vehicles. Under the condition (13) free flow remains at the bottleneck. If in contrast,

$$\gamma > \gamma_{\text{cr}}, \tag{14}$$

then the ACC-vehicles *induce* the onset of congestion at the bottleneck where free flow without ACC-vehicles occurs (Fig. 7 (b)). As a result of the onset of congestion, a GP appears at the bottleneck (Fig. 7 (b)). The higher the percentage of the ACC-vehicles $\gamma$, the more frequency of the moving jam emergence in the GP (Fig. 7 (c, d)).

We can see from these two examples of the ACC-vehicles that the ACC-vehicles can influence congested pattern features qualitatively. However this occurs only if the percentage of the ACC-vehicles $\gamma > \gamma_{\text{cr}}$. The critical value $\gamma_{\text{cr}}$ strong depends on the ACC-vehicle dynamics. It can be assumed that if the ACC-vehicle possesses qualitative different dynamical features than the one described by the equation of the ACC-dynamics (12), then quantitative different results of the ACC-vehicle influence on congested patterns at freeway bottlenecks can be expected. Thus, there is a great potential for the development of new vehicle assistance systems which can improve traffic comfort and increase traffic safety considerably due to their influence on congested pattern features.

**CONCLUSIONS**

The investigation made above allows us to conclude the following:

(i) To find the efficiency of traffic control and management strategies, an influence of these strategies on dynamical *spatial-temporal* features of congested patterns at freeway bottlenecks should first be studied.

(ii) The on-ramp metering control strategy with a feedback is an efficient method for an increase in the flow rate on a freeway and for a decrease in traffic congestion at bottlenecks.

(iii) Changes in the ramp inflow at the bottleneck where a congested pattern has occurred and in the flow rate upstream of the pattern can lead to qualitative different consequences for the spatial-temporal pattern dynamics and the pattern dissolution.

(iv) The traffic control based on autonomous ACC functions in vehicles can improve the efficiency of the system: Wide moving jams can be suppressed by ACC-vehicles and therefore traffic flow can be harmonized and stabilized in a synchronized mode.

(v) At certain parameters and percentages of ACC traffic flow can be influenced in a negative direction: ACC function can possibly lead to traffic breakdowns and the congested pattern occurrence at bottlenecks.

The latter two results support the importance of the microscopic modeling based on the three-phase traffic theory: Different kinds of vehicle assistance systems can be investigated in simulations concerning their influence on the efficiency of the traffic control system. In particular, this is related to such traffic assistance systems are "jam warning", "improved acceleration after wide moving jam", "stop-and-go assistance" and so on.

(vi) The microscopic simulation environment based on the three-phase traffic theory is in a very good accordance with empirical *spatial-temporal* features of congested patterns. Therefore this is a high value instrument in the investigation of different freeway traffic control strategies as well of traffic assistance systems before they are introduced in the market.

**ACKNOWLEDGEMENT**

I thank Sergey Klenov for his help and the German Ministry of Education and Research for financial support within the BMBF project "DAISY".

Figure captions

Fig. 1. Steady state model solutions (a) *(17)*, the diagram of congested patterns at the bottleneck due to the on-ramp (b) *(14), (17)*, the schema of the bottleneck due to the on-ramp (c).

Fig. 2. Congested pattern under the on-ramp inflow control. $q_{in}$ =1945, $q_{on}$ =350, $\bar{q}_{on}$ =227 (b), 235 (c), 243 (d) vehicles/h. The on-ramp inflow is switched on at $t = t_0$ =8 min. In (g, h) 1-min averaged data.

Fig. 3. Dissolving of GP: (b-f) – Variation of the incoming flow rate $q_{in}$. (g, h) – Variation of the on-ramp flow rate $q_{on}$ (g), and variation of both $q_{in}$ and $q_{on}$ (h).

Fig. 4. Dissolving of WSP. Variation of the incoming flow rate $q_{in}$. $q_{in}$ are 2230 (a), 2120 (b), 2000 (c), 1950 (c) vehicles/h. $q_{on}$ =190 vehicles/h.

Fig. 5. Initial congested patterns (left) (taken from *(18)*) and their dissolving (right) on the road with the upstream ('U') and the downstream ('D') on-ramps due to a variation of the on-ramp flow rates.

Fig. 6. Influence of the ACC-vehicles with $K_1 = 0.1$, $K_2 = 0.55$ on GP at the on-ramp. $q_{on} = 600$, $q_{in} = 1730$ vehicles/h. The percentage of the ACC-vehicles $\gamma$ is 0% (a), 20% (b), 30% (c), 37% (d).

Fig. 7. Influence of the ACC-vehicles with $K_1 = 0.03$, $K_2 = 0.18$ on the onset of congestion at the bottleneck due to the on-ramp. $q_{on} = 400$, $q_{in} = 1730$ vehicles/h. $\gamma$ is 0% (a), 5% (b), 8% (c), 20% (d).

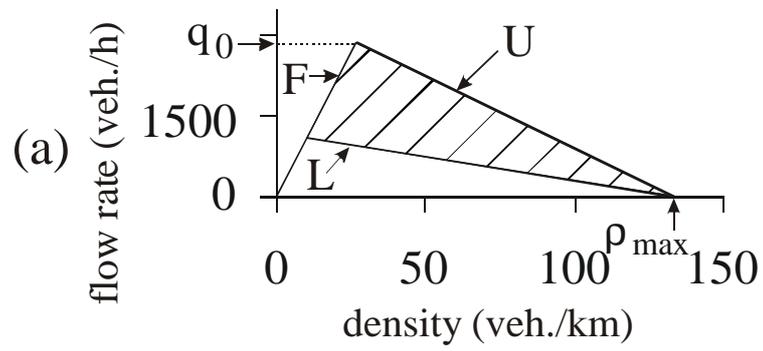

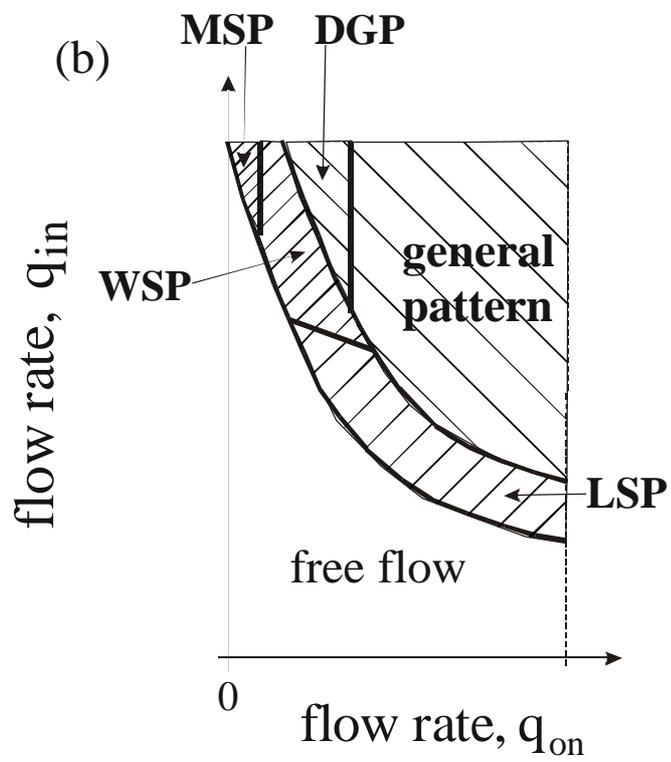

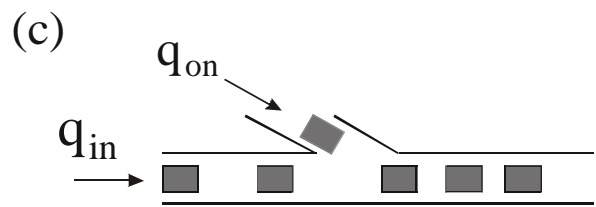

Fig. 1

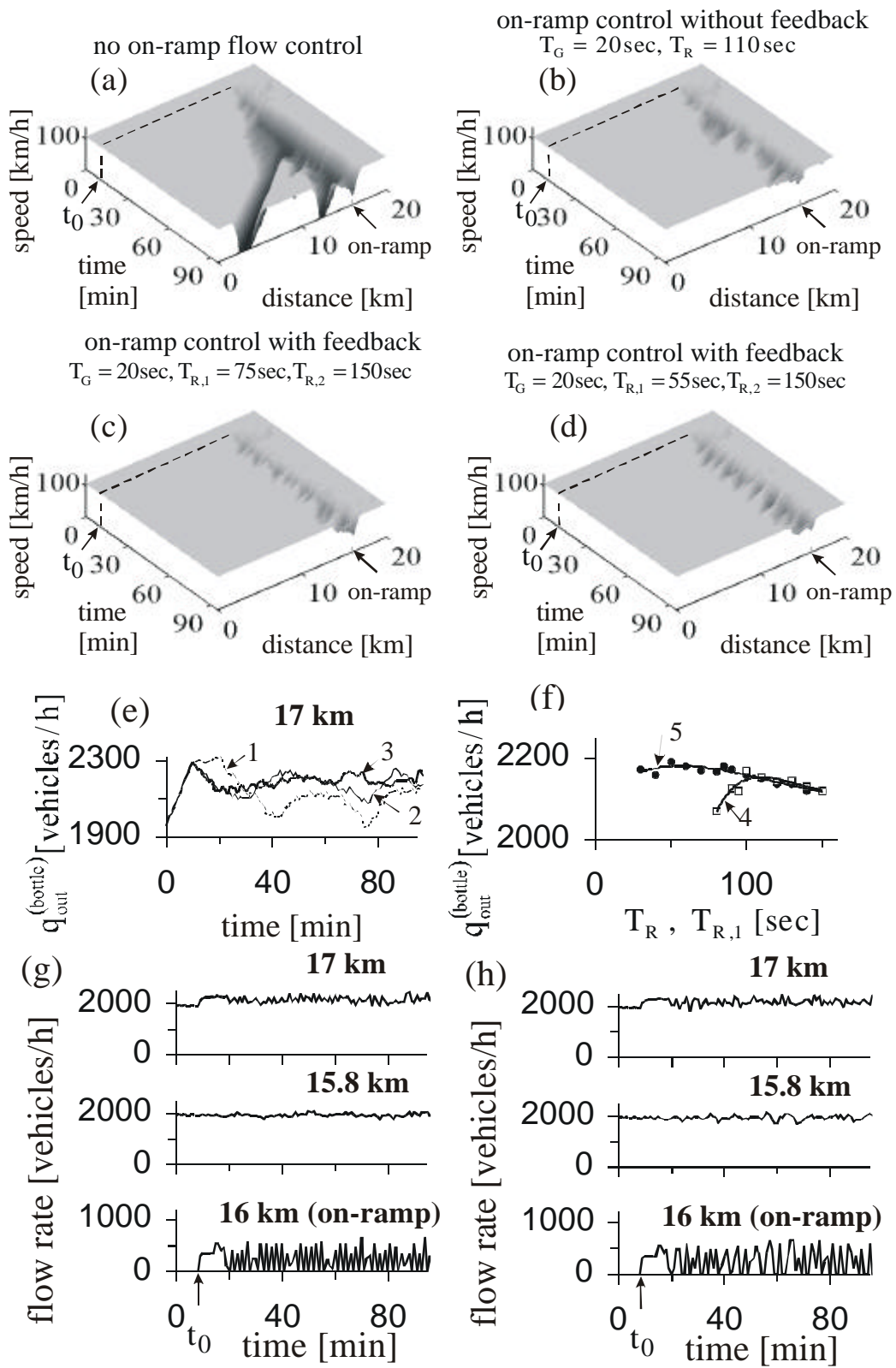

Fig. 2

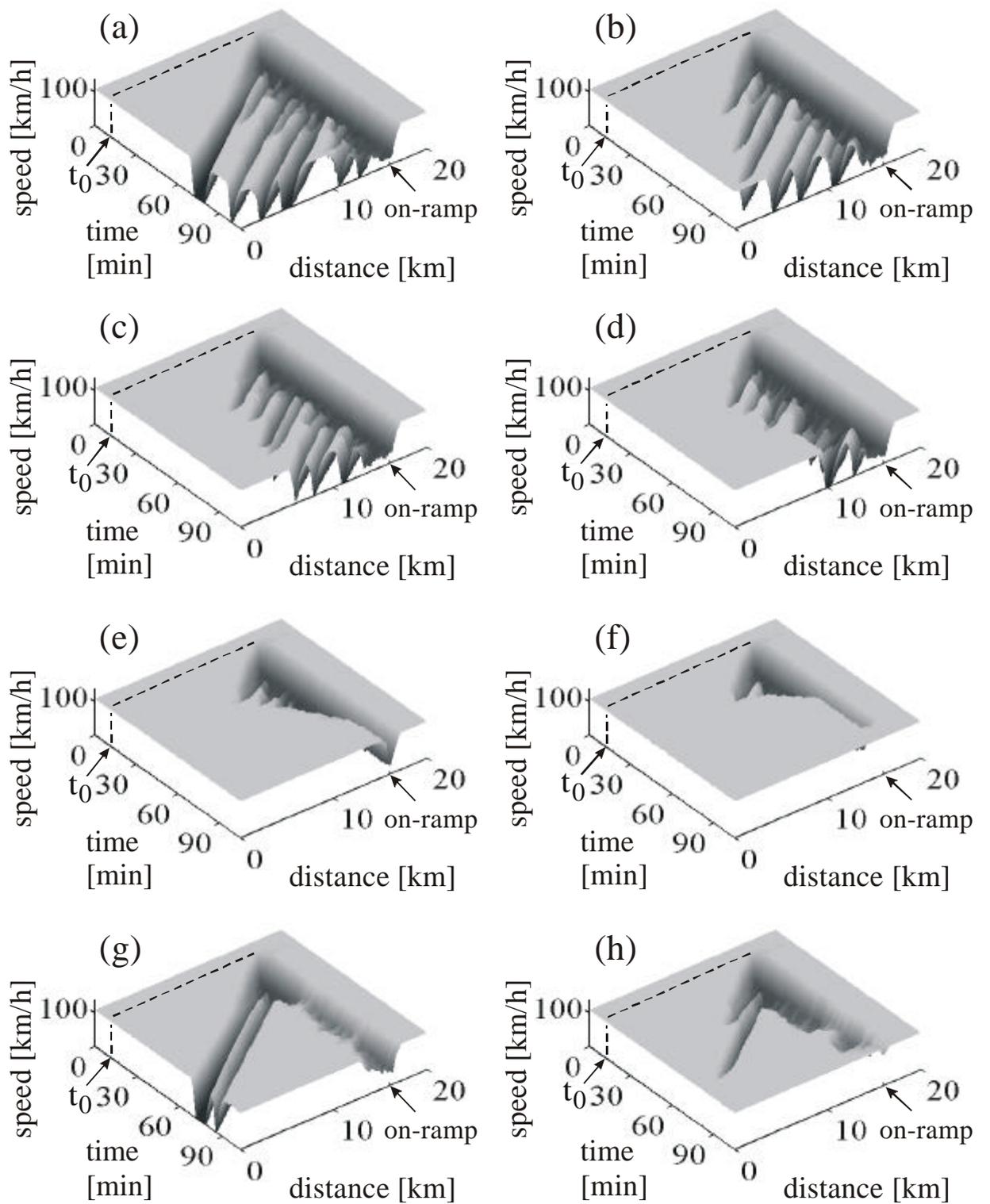

Fig. 3

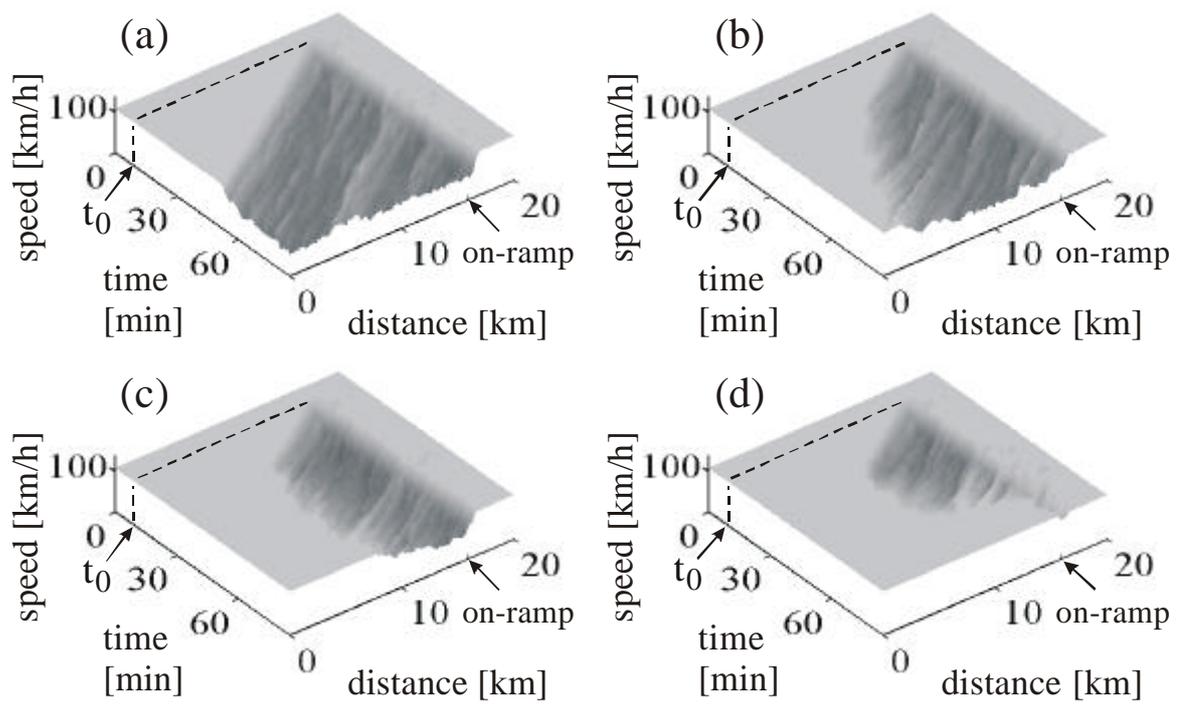

Fig. 4

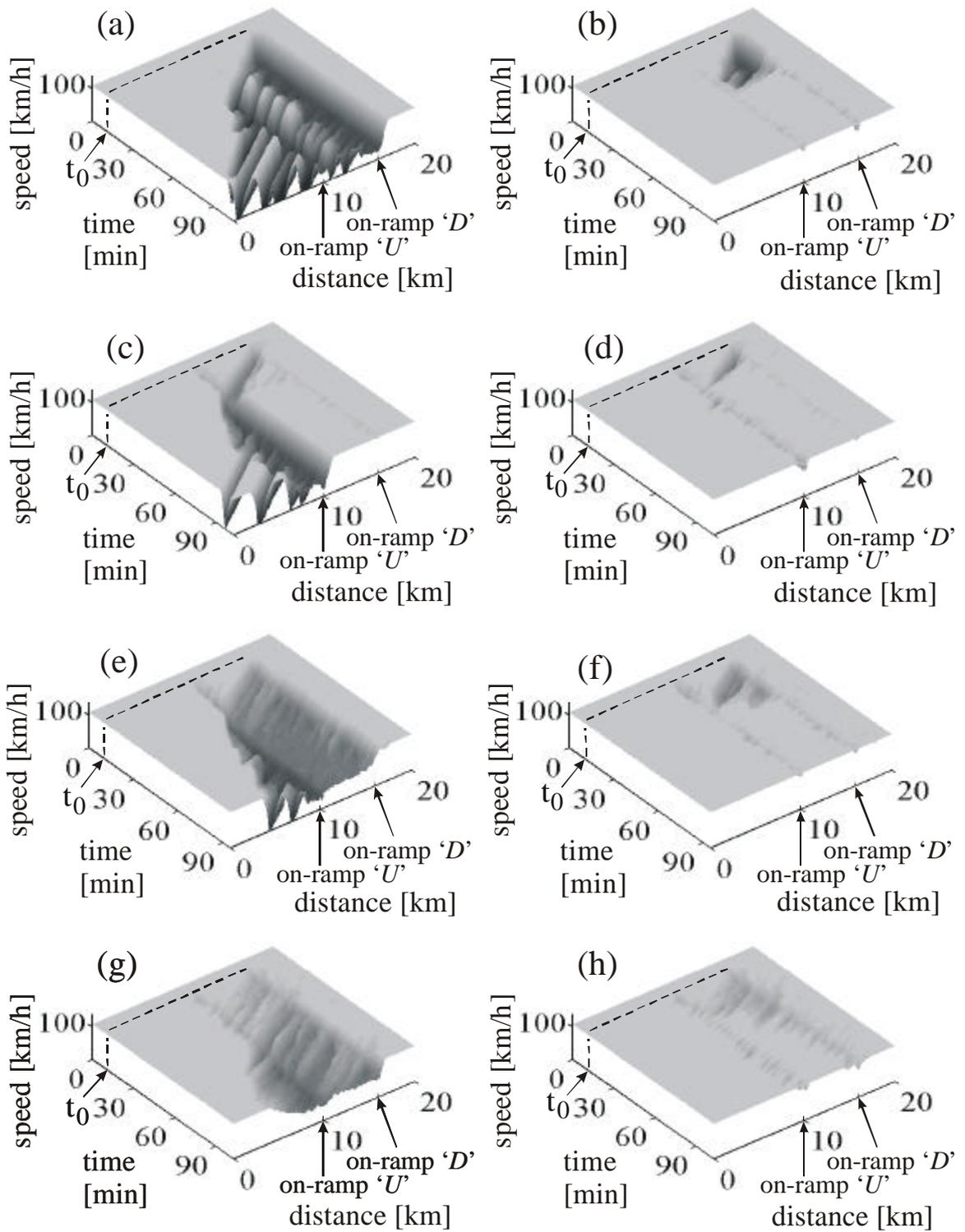

Fig. 5

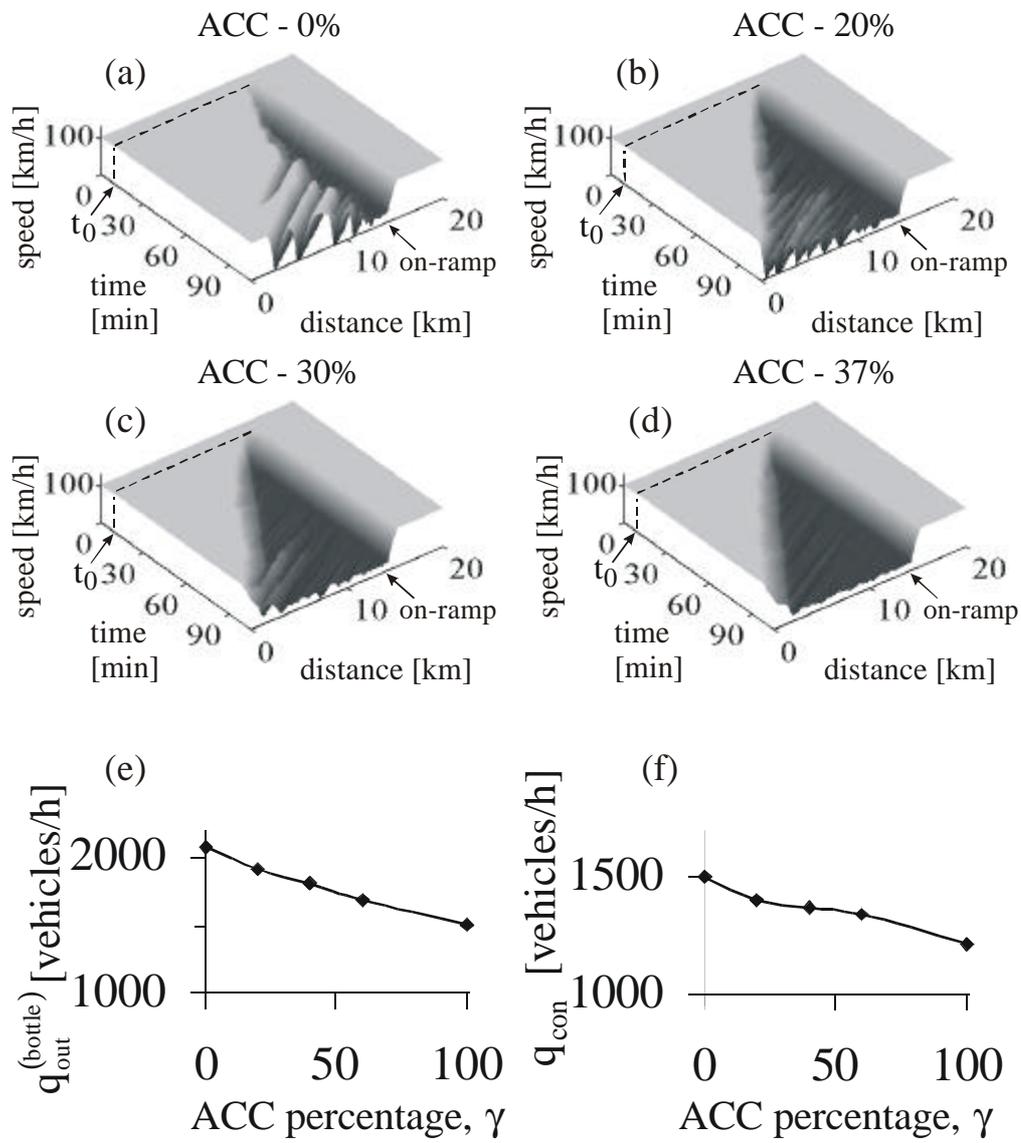

Fig. 6

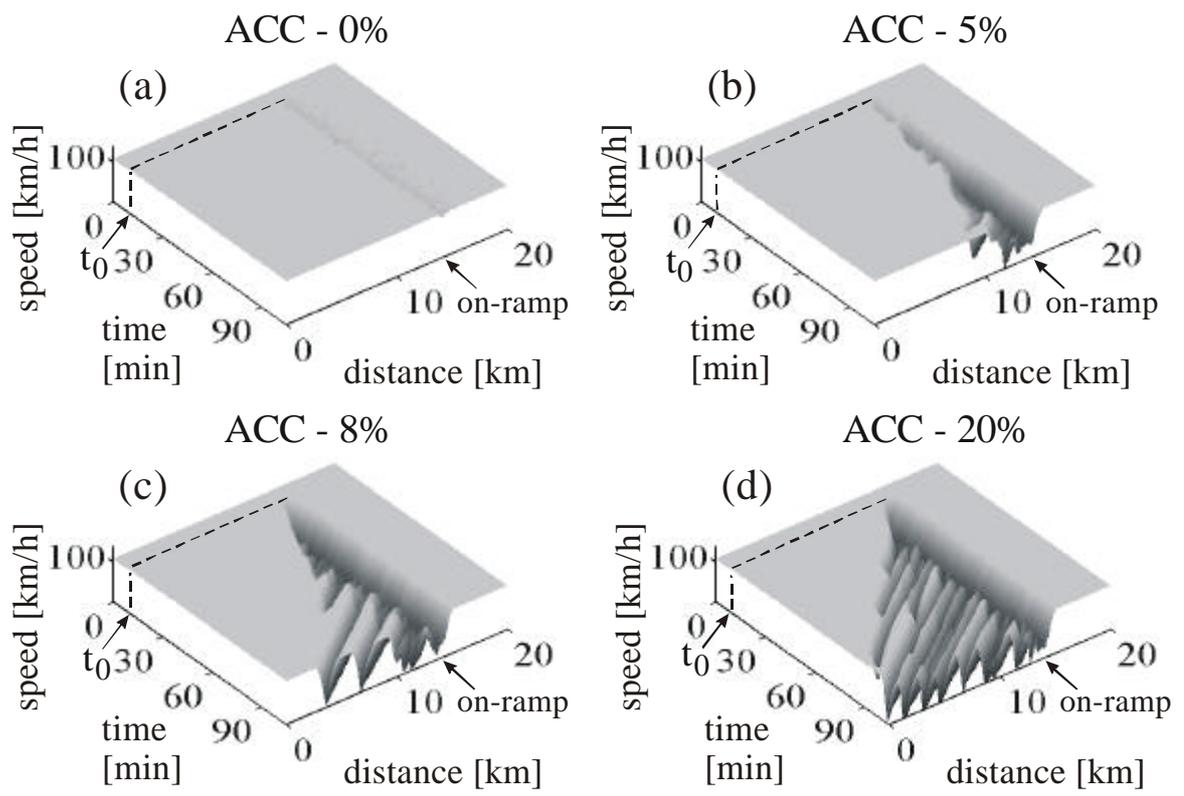

Fig. 7